\documentclass[preprint,proceedings]{rmaa}

\usepackage{paralist}

\usepackage{psfrag,color}

 \def\microns{\mbox{ } \mu \mbox{m}}
\def\deg{^{\circ}}
 
\SetYear{2003}
\SetConfTitle{Compact Binaries in the Galaxy and beyond}

\title{Revealing the nature of the highly obscured galactic source IGR J16318$-$4848\altaffilmark{0}} 

\author{
  S. Chaty,\altaffilmark{1,2,4} 
  and P. Filliatre\altaffilmark{2,3}
  }

\altaffiltext{1}{Universit\'e Paris 7, 2 place Jussieu,
F-75 005 Paris, France.}
\altaffiltext{2}{Service d'Astrophysique, DSM/DAPNIA/SAp,
CEA/Saclay, F-91 191 Gif-sur-Yvette, Cedex, France.}
\altaffiltext{3}{F\'ed\'eration de Recherche APC, Paris, France.}
\altaffiltext{4}{chaty@cea.fr}
\altaffiltext{0}{Based on observations collected at the European Southern Observatory, Chile (observing proposal ESO N$^o$ 70.D-0340).}

\shortauthor{Chaty \& Filliatre}

\suppressfulladdresses

\listofauthors{S. Chaty \& P. Filliatre}
\indexauthor{Chaty, S.}
\indexauthor{Filliatre, P.}

\abstract{The X-ray source IGR J16318$-$4848 was the first source discovered by INTEGRAL on 2003, January 29. 
We carried out optical and near-infrared (NIR) observations 
at the European Southern Observatory (ESO La Silla) in the course of a Target of Opportunity (ToO) programme. 
We discovered the optical counterpart and confirmed an already proposed NIR
candidate.
NIR spectroscopy revealed a large amount of emission lines, including forbidden iron lines and P-Cygni profiles. The spectral energy distribution of the source points towards a high luminosity and a high temperature,
 with an absorption greater than the interstellar absorption, but two orders of magnitude lower than the X-ray absorption. We show that the source is an High Mass X-ray binary (HMXB) at a distance between $\sim 1$ and $\sim 6$ kpc, 
the mass donor being an early-type star, probably a sgB[e] star, surrounded by a rich and absorbing circumstellar material. This would make the second High Mass X-ray Binary (HMXB) with a sgB[e] star after CI Cam, indicating that a new class of strongly absorbed X-ray binaries is being unveiled by INTEGRAL.
}

\addkeyword{stars: circumstellar matter}
\addkeyword{stars: emission-line}
\addkeyword{Be---X-rays: binaries}
\addkeyword{IGR J16318-4848}

\begin{document}
\maketitle

\begin{figure}[!t]
  \includegraphics[width=\columnwidth]{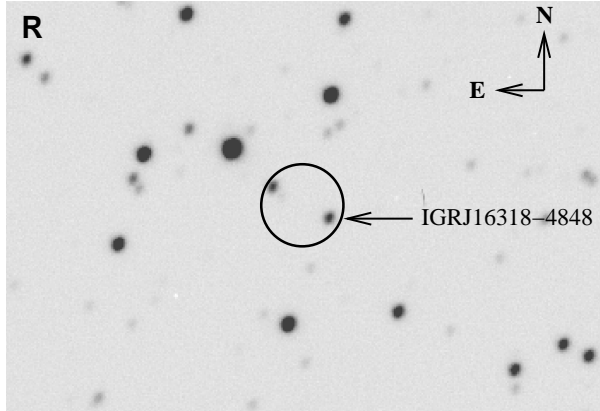}
  \caption{R band image of the field of view of IGR J16318$-$4848.
We reported the XMM uncertainty circle of $4\arcsec$.}
  \label{fig:f4}
\end{figure}
\begin{figure}[!t]
  \includegraphics[width=\columnwidth]{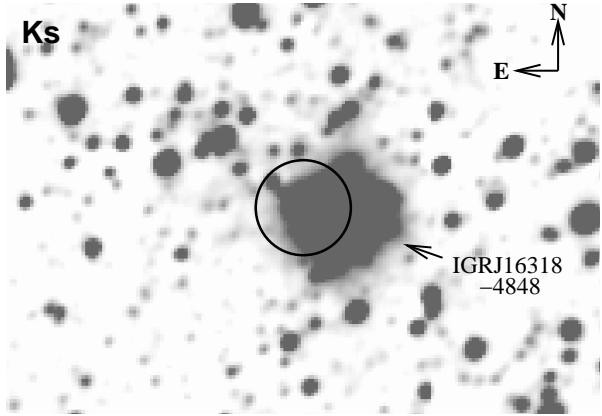}
  \caption{Ks band image of the same field.}
  \label{fig:f6}
\end{figure}
\begin{figure}[!t]
  \includegraphics[angle=0,width=\columnwidth]{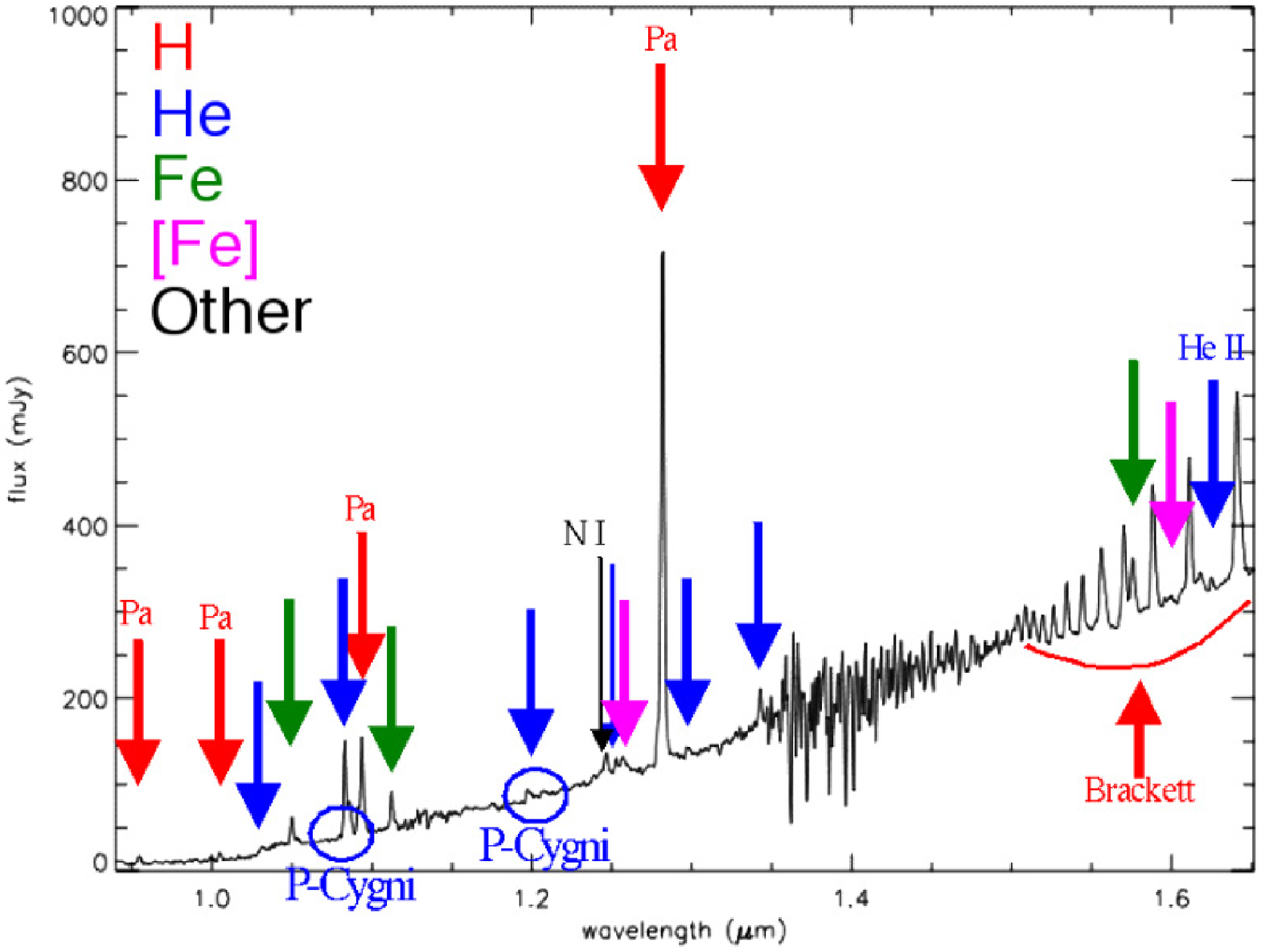}
  \caption{NIR spectrum (0.95$-$1.65 $\microns$)}
  \label{fig:f15}
\end{figure}
\begin{figure}[!t]
  \includegraphics[angle=0,width=\columnwidth]{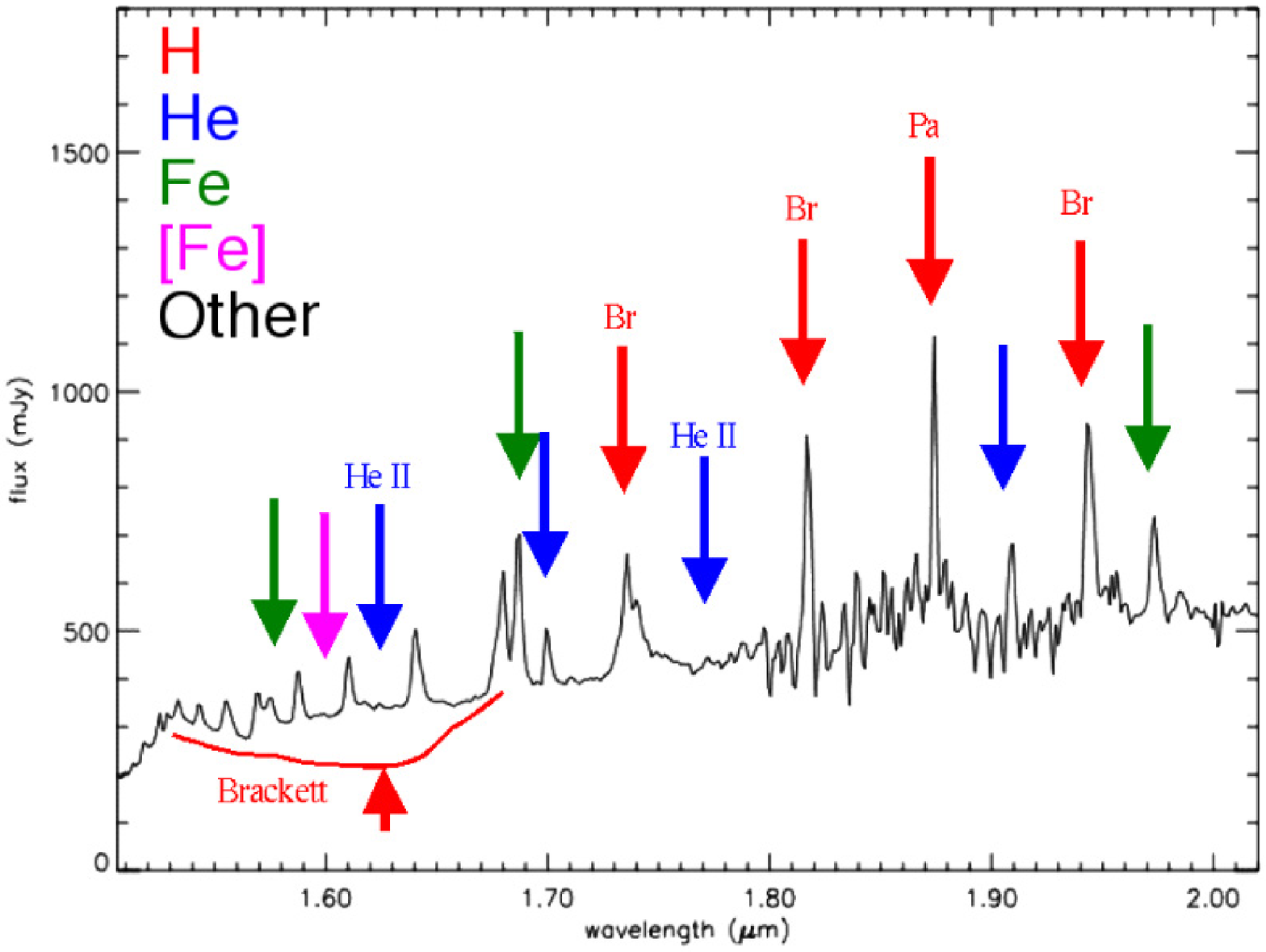}
  \caption{NIR spectrum (1.5$-$2.05 $\microns$)}
  \label{fig:f16}
\end{figure}
\begin{figure}[!t]
  \includegraphics[angle=0,width=\columnwidth]{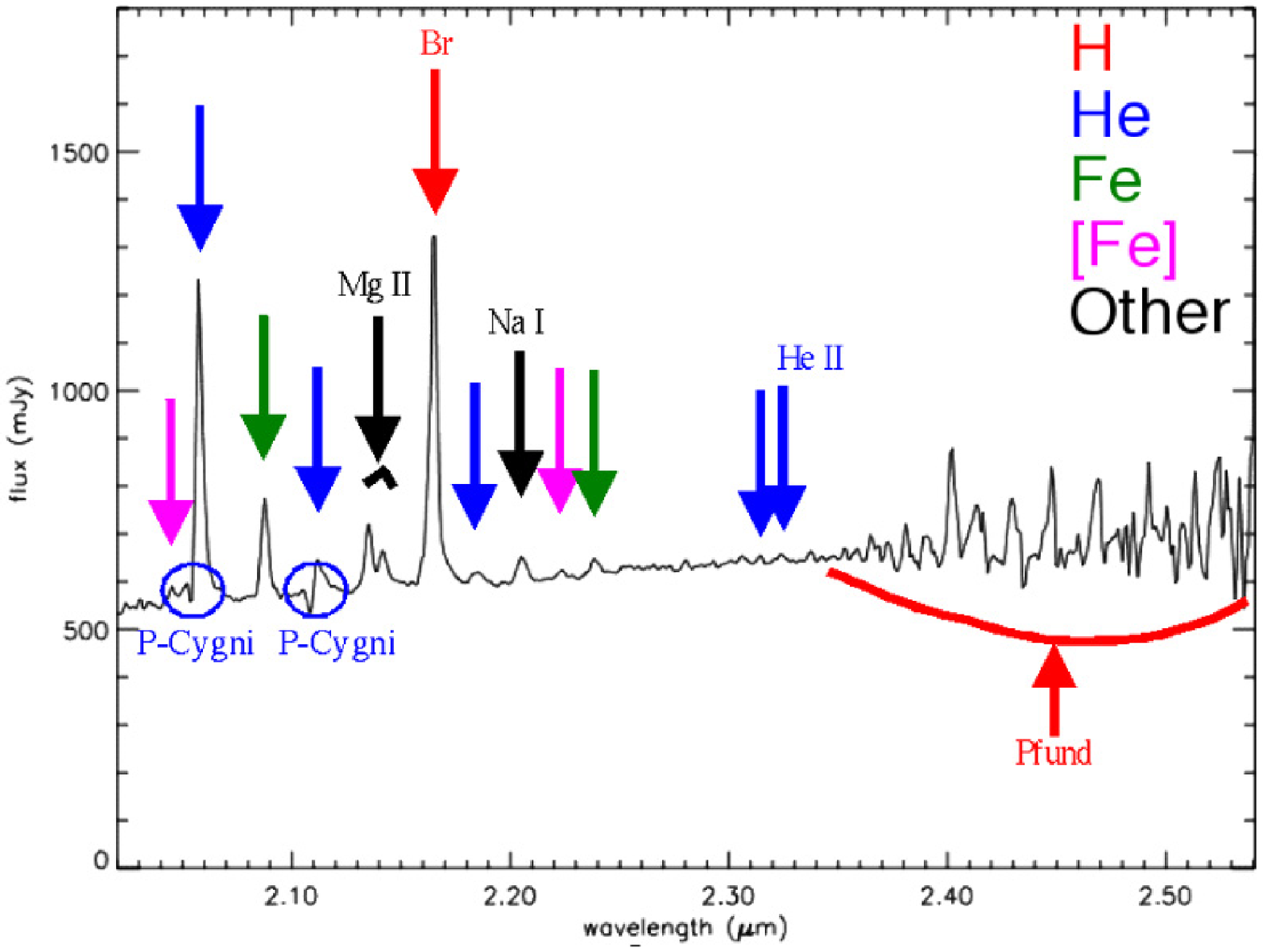}
  \caption{NIR spectrum (2.0$-$2.55 $\microns$)}
  \label{fig:f17}
\end{figure}

IGR J16318$-$4848 has been the first source to be discovered by 
the INTEGRAL imager IBIS/ISGRI on 2003 January 29, 
$0.5\deg$ south from the galactic equator \citep{courvoisier}.
A subsequent observation by XMM-Newton 
on 2003 February 10 localized it with an accuracy of $4\arcsec$,
the high energy spectrum suggesting
a high column density of $N_{\rm H}> 10^{24}\,\rm cm^{-2}$
(\citet{matt}, \citet{walter}).
In the course of a ToO programme at ESO to look for counterparts of
 high energy sources discovered by satellites including INTEGRAL (PI S. Chaty), we carried out on 2003, February, 23-25 photometric and spectroscopic observations in the optical and NIR
of the high-energy source IGR J16318$-$4848, with EMMI and SOFI instruments on ESO/NTT. 
We discovered the optical counterpart and confirmed the NIR one 
(see \citet{walter}) by independent astrometry.
The optical/NIR images and spectra are shown in Figures \ref{fig:f4},
\ref{fig:f6}, \ref{fig:f15}, \ref{fig:f16} and \ref{fig:f17}.
We derived the absorption along the line of sight: Av$\sim 17.4$ magnitudes,
and the temperature of the companion star: $\sim 18000$ K.
The distance of the source is constrained between $\sim 1$ and $\sim 6$ kpc.
The $0.95 - 2.52\,\micron$ NIR spectrum is highly unusual, very rich in emission lines, suggesting 
a highly complex and stratified circumstellar environment, or an enveloppe.
Study of the spectral lines suggest a sgB[e] star so the system
would be a HMXB, probably hosting a neutron star,
like CI Cam.
The reader should consult \citet{filliatre} for more details.
INTEGRAL is on the course of revealing a new population of
obscured high energy sources, which might help us to understand 
the evolution of high-energy binary systems.

\acknowledgements

SC is grateful to Rob Hynes, Joanna Mikolajewska, Ignacio Negueruela 
and Marc Rib\'o for very helpful
discussions on the nature of the source during this IAU colloquium.

\end{document}